\documentclass[12pt]{article}
\usepackage{a4wide}
\usepackage{amsmath}
\usepackage{amssymb}
\usepackage{graphicx}
\usepackage{latexsym,epsfig,cite}
\usepackage{booktabs}
\usepackage{hyperref}

\newcommand{\bTheta}{\mathbf{\Theta}}
\newcommand{\bD}{\mathbf{D}}
\newcommand{\llh}{\mathcal{L}}
\newcommand{\ev}{\mathcal{Z}}

\def\beq{\begin{equation}}
\def\eeq{\end{equation}}
\def\bea{\begin{eqnarray}}
\def\eea{\end{eqnarray}}

\def\lsim{\mathrel{\rlap{\raise 2.5pt \hbox{$<$}}\lower 2.5pt\hbox{$\sim$}}}
\def\gsim{\mathrel{\rlap{\raise 2.5pt \hbox{$>$}}\lower 2.5pt\hbox{$\sim$}}}

\allowdisplaybreaks 

\usepackage[dvips]{color}
\definecolor{Black}{named}{Black}
\definecolor{Red}{named}{Red}

\begin{document}

\thispagestyle{empty}
\begin{flushright}

\end{flushright}
\vspace*{5mm}
\renewcommand*{\thefootnote}{\fnsymbol{footnote}}
\setcounter{footnote}{3}

\begin{center}
{\large {\bf Long lived charginos in Natural SUSY?
}}\\
\vspace*{1cm}
{\bf N.-E.\ Bomark$^1$\footnote{From 01.09.2013 at National Centre for Nuclear Research, Ho\.za 69, 00-681 Warsaw, Poland.}, A.\ Kvellestad$^2$, S.\ Lola$^3$,}
{\bf P.\ Osland$^1$ and A.R.\ Raklev$^2$} \\
\vspace{0.3cm}
$^1$ Department of Physics and Technology, University of Bergen,
N-5020 Bergen, Norway\\
$^2$ Department of Physics,
University of Oslo, N-0316 Oslo, Norway\\
$^3$ Department of Physics, University of Patras, GR-26500 Patras, Greece
\end{center}

\begin{abstract}
Supersymmetric models with a small chargino--neutralino mass
difference, and as a result metastable charginos, have been a popular
topic of investigation in collider phenomenology. The possibility of a chargino lighter than the lightest neutralino has also been raised.
Recently, the absence of any supersymmetric signal in the 8 TeV LHC data has led to
significant interest in
the so-called Natural SUSY models with light higgsinos. These models also have a naturally small chargino--neutralino mass difference.  However, we show here that when relevant indirect constraints from results at the LHC and elsewhere are applied, this possibility is heavily constrained within the Minimal Supersymmetric Standard Model (MSSM): massive metastable higgsinos are not a signature of Natural SUSY.
\end{abstract}

\renewcommand*{\thefootnote}{\arabic{footnote}}
\setcounter{footnote}{0}
\setcounter{page}{1}

\section{Introduction}

Models with near-degenerate electroweak gauginos have long been studied in
the context of anomaly-mediated breaking of supersymmetry
(SUSY)~\cite{Randall:1998uk,Giudice:1998xp}. The absence of any SUSY
signal at the LHC and the question of the naturalness of the theory
has recently led to the consideration of the so-called Natural SUSY
models, where only the higgsinos, the stops, the left-handed sbottom,
and, to a more limited extent, the gluino, are necessarily light
enough to be probed at the
LHC~\cite{Brust:2011tb,Papucci:2011wy}. Higgsino dominance of light
neutralinos and charginos in these models could also lead to a small
mass difference between the two particles. Such spectra would in turn
result in rather characteristic experimental signals in high-energy
collisions, including charginos that live long enough to create
displaced vertices, or even pass through the detector before decaying,
if the mass difference is sufficiently small
\cite{Chen:1996ap,Feng:1999fu,Gherghetta:1999sw}. Considerable effort
has been invested in searching for such massive metastable charged
particles (MMCPs), see {\it e.g.}\ the reviews
\cite{Fairbairn:2006gg,Raklev:2009mg}, and the most recent experimental limits~\cite{ATLAS-CONF-2013-058,Aad:2013yna,Chatrchyan:2013oca}.

This mass difference is also important to searches for Natural SUSY through the production of stop quarks. The decay $\tilde{t}_1 \to t\tilde\chi_1^0$ will have competition from a significant fraction of  $\tilde{t}_1 \to b\tilde\chi^+_1$ decays, where a small mass difference between the resulting chargino and the neutralino will make the decay products, other than the $b$--quark, very soft and difficult to detect.

As discussed by Kribs {\it et al.}\ \cite{Kribs:2008hq}, in the
general Minimal Supersymmetric Standard Model (MSSM) the chargino
could in principle be even lighter than the lightest neutralino in a
corner of the parameter space, with the gravitino (or axino)
constituting dark matter, thus avoiding a completely stable massive charged particle. Note that this is not in line with the standard anomaly-mediated breaking scenario, where the gravitino is heavy compared to the other sparticles which have loop-suppressed masses.

In this paper, we investigate how degenerate the lightest neutralino
and chargino can be in Natural SUSY models based on the MSSM.  We
explore to what extent these models are compatible with current
collider constraints, and in particular with the discovery of a new boson at the LHC, when interpreted as the light SM-like Higgs state of the MSSM~\cite{Aad:2012tfa,Chatrchyan:2012ufa}. This is done by scanning the relevant parameter space using the Bayesian inference tool {\tt MultiNest~2.17}~\cite{Feroz:2007kg,Feroz:2008xx} for the analysis of posterior probability distributions. We find that there is very little room left for degeneracy when requiring that the models are compatible with existing constraints, in particular for a higgsino chargino--neutralino pair, and that the possibility of negative mass differences discussed in \cite{Kribs:2008hq} is effectively ruled out in Natural SUSY.

We begin in Section \ref{sec:massdiff} by discussing the parameters
that affect the chargino--neutralino mass difference in
the MSSM and the parameter space of Natural SUSY. The scan that we
have performed is described in
Section~\ref{sec:scan}. Finally we describe the consequences of our results in Section~\ref{sec:imp}, discussing in particular the implications for chargino decay length, before we conclude in Section~\ref{sec:conclusions}.

\section{The relevant parameter space of Natural SUSY}
\label{sec:massdiff}
\subsection{Chargino--neutralino mass difference}
\label{sec:mdiff}

In the MSSM, the free mass parameters of the neutralino and chargino mass matrices at tree level are $M_1$, $M_2$ and $\mu$.\footnote{For the model parameters we adopt the notation of Martin~\cite{Martin:1997ns}.} In addition $\tan\beta$ also enters as a free parameter. The complex phases of the mass parameters are very constrained, in particular,  due to the electric dipole moments \cite{Feng:2008cn,Ellis:2008zy,Cheung:2009fc,Altmannshofer:2009ne}. However, there are no {\it a priori} grounds not to give arbitrary signs to these, although by a rotation of basis we can choose $M_2$ to always be positive.

For small $M_1$ the lightest neutralino will be a bino, which is
historically the most popular choice. Here there is no degeneracy
between the chargino and the neutralino, because $M_1$ does not enter into the chargino mass matrix.  When $\mu$ or $M_2$ is the smallest parameter we may have a higgsino or wino neutralino, and in both cases there may be degeneracy with the chargino.

In Natural SUSY models the higgsino limit is commonly realised.\footnote{Note that, as we will see later, a substantial bino or wino component in the lightest neutralino is not {\it a priori} impossible in Natural SUSY, however, large values for the parameters $M_1$ and $M_2$ do not heavily penalise the naturalness of the model.} Here, with $M_2>\mu,M_W$, the tree-level mass difference $\Delta m\equiv m_{\tilde\chi_1^\pm}-m_{\tilde\chi_1^0}$ from an expansion in $1/M_2$ is \cite{Giudice:1995qk}
\begin{equation}
\Delta m = \left[\frac{M_2}{M_1}\tan^2\theta_W+1+{\rm sgn}\,\mu\left(\frac{M_2}{M_1}\tan^2\theta_W-1\right)\sin{2\beta}\right]\frac{M_W^2}{2M_2}+{\mathcal O}\left(\frac{1}{M_2^2}\right).\label{eq:dmhiggsino}
\end{equation}
For positive $M_1$ and $M_2$, $\Delta m$ is in this limit always positive. It gets small for very large $M_1,M_2\gg M_W$, but numerically this does not decrease the mass difference below 300 MeV unless both masses are larger than ${\mathcal O}(10~{\rm TeV})$, or if $\tan\beta\simeq 1$ and either mass is very large.
The above expansion breaks down if $\mu\to 0$, however, because of LEP bounds on the $Z$--decay, we know that $\mu>M_Z/2$ as the chargino would otherwise contribute in the decay.

\begin{figure}
\begin{center}
\includegraphics[width=0.45\textwidth]{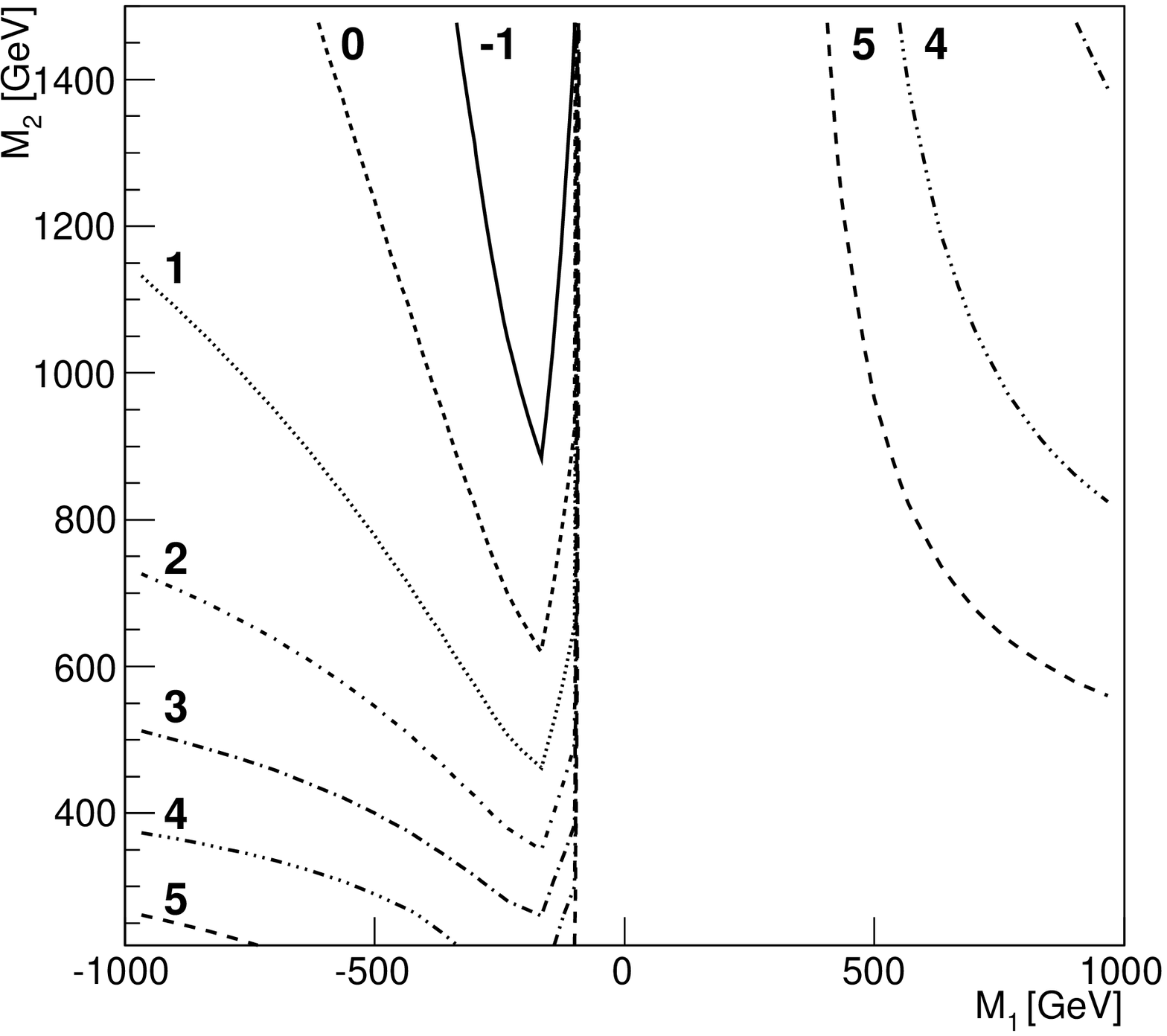}
\includegraphics[width=0.45\textwidth]{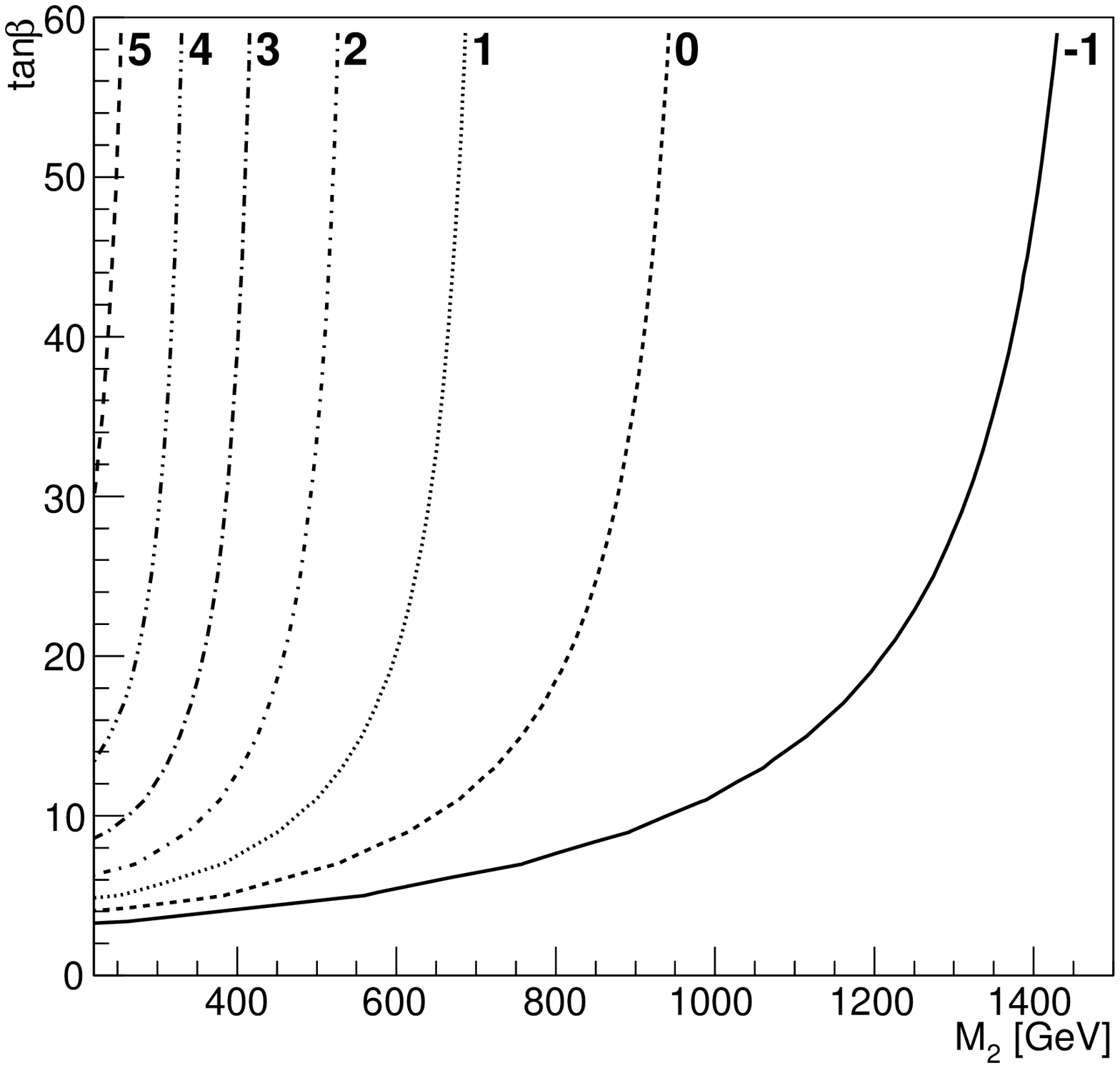}
\end{center}
\caption{Contours of mass difference $\Delta m$ for $\tan\beta = 10$ and $\mu = 80$\;GeV (left), and $M_1 = -200$\;GeV and  $\mu = 80$\;GeV (right).}\label{fig:mdiff}
\end{figure}

However, $\Delta m >0$ is no longer necessarily the case for negative values of $M_1$. In Fig.~\ref{fig:mdiff} we show contours for the mass difference
$\Delta m$ (using Eq.~(\ref{eq:dmhiggsino}) expanded to include terms of order $1/M_2^2$), in the $M_1-M_2$ mass plane and in the $M_2-\tan\beta$ plane. This demonstrates that very small and even  negative mass differences are possible at tree level, for negative $M_1$ and for relatively large values of $M_2$. Larger values of $|\mu|$, and negative $\mu$, require larger values of $M_2$ for this to happen. We also see that small and negative mass differences favour low $\tan\beta$.

The leading higher order corrections to Eq.~(\ref{eq:dmhiggsino}) stem from top--stop loops and $\gamma(Z)$--higgsino loops. These are generally small unless $\tan\beta$ is small and/or the stop mixing is near maximal and/or for large values of $|\mu|$. In the scans that follow these effects are included.

In the wino limit, $M_2<|\mu|,|M_1|$, the corresponding expression for the mass difference is
\begin{eqnarray}
\Delta m &=& \frac{M_W^2}{\mu^2}\frac{M_W^2}{M_1-M_2}\tan^2\theta_W\sin^2{2\beta}+2\frac{M_W^4M_2\sin{2\beta}}{(M_1-M_2)\mu^3}\tan^2\theta_W\nonumber\\
&&+\frac{M_W^6\sin^3{2\beta}}{(M_1-M_2)^2\mu^3}\tan^2\theta_W(\tan^2\theta_W-1)+{\mathcal O}\left(\frac{1}{\mu^4}\right),
\label{eq:dmwinotree}
\end{eqnarray}
when expanded in $1/\mu$~\cite{Feng:1999fu,Gherghetta:1999sw}. This is small for large values of $\tan\beta$, and in the limit of $\tan\beta\to\infty$ the lowest contributing order is $1/\mu^4$. In this case the mass difference is dominated by loop corrections.

\subsection{Natural SUSY}

The essential phenomenological properties of Natural SUSY are the existence of two light stop quarks, a light left-handed sbottom quark, a light higgsino chargino and neutralino, and a relatively light gluino. These are the necessary ingredients that provide a less fine-tuned electroweak sector.

In order to cover all the MSSM parameter space that features Natural SUSY we must investigate a suitable subset of the MSSM parameters. For EWSB we choose $\mu$, $m_{A^0}$ and $\tan\beta$ as the free parameters, and derive $m_{H_u}$ and $m_{H_d}$. For the higgsinos we further need the gaugino masses $M_1$ and $M_2$. The gluino mass is also a free parameter through $M_3$. The properties of the light squarks are covered by separate soft masses for the third generation squarks, $m_{Q_3}$, $m_{u_3}$ and  $m_{d_3}$ and a free choice of trilinear term for the squark mixing $A_0$.
For the remaining MSSM parameters we assume that the squarks and sleptons are too massive to be relevant, and we simply fix a common high value of 3 TeV for their corresponding soft masses. All parameters are defined at a scale of 1 TeV, except for $\tan\beta$ and the pole mass $m_{A^0}$. In total we have ten free parameters for our model of Natural SUSY. The question of what priors and parameter ranges to use for the scan will be discussed in Section~\ref{sec:scan_set-up}.



\section{Parameter scan}
\label{sec:scan}

\subsection{Bayesian parameter estimation}
\label{sec:bayes}

In order to study the parameter space  of Natural SUSY we employ Bayesian parameter estimation. In this section we review the basic concepts of this method most relevant for our analysis.

Starting from a model $H$ with a set of parameters $\bTheta$, such as our parameterisation of Natural SUSY, the goal of Bayesian parameter estimation is to use a set of data $\bD$ to determine the posterior probability distribution for the parameters, $P(\bTheta|\bD,H)$.\footnote{Here we recall that probability in the Bayesian sense represents a \textit{degree of belief} in a hypothesis or a choice of parameter values.} From Bayes' theorem,
\begin{equation}
  P(\bTheta|\bD,H) = \frac{P(\bD|\bTheta,H) P(\bTheta|H)}{P(\bD|H)}
                   = \frac{\llh(\bTheta) \pi(\bTheta)}{\ev},
\end{equation}
the posterior distribution is given in terms of the \textit{likelihood}  $\llh(\bTheta)=P(\bD|\bTheta,H) $ and the \textit{prior} belief in the values of the parameters of the model $ \pi(\bTheta)=P(\bTheta|H)$, with the \textit{Bayesian evidence} $ \ev=P(\bD|H) $ as a normalization factor.

For any parameter point $\bTheta$ in the model, a probability density function (PDF) for the outcome of an experiment can be constructed, {\it e.g.}\ a Gaussian centered at the outcome predicted by the given model parameter point.
The likelihood function is then obtained by evaluating this PDF for the case where the outcome equals the observed data $\bD$ and interpreting the resulting expression as a function of the model parameters.
Thus the likelihood quantifies the level of agreement between the model and experiment across the parameter space. 

The prior is the PDF describing our degree of belief in a given parameter point before confronting the model with the data, {\it e.g.}\ on the basis of previous experiments or theoretical consideration. The particular priors and likelihoods used will be discussed in the next section; for reference they are to be found in Tables~\ref{tab:ScanPar} and~\ref{tab:Constraints}, respectively. Finally, the Bayesian evidence is given by
\begin{equation}
  \ev = \int \llh(\bTheta) \pi(\bTheta)\, d\bTheta.
  \label{eq:evidence}
\end{equation}
The evidence is a key quantity when doing Bayesian model comparison, but for the purpose of parameter estimation it only serves to normalize the posterior distribution.


For models with a large number of parameters and a complicated mapping from parameters to observables, the posterior $P(\bTheta|\bD,H)$ can be approximated by Monte Carlo methods. The result is a set of parameter points $\bTheta_i$ distributed according to $P(\bTheta|\bD,H)$. Here we make use of the {\tt MultiNest} algorithm, to which we supply a likelihood $\llh(\bTheta)$ and a prior $\pi(\bTheta)$. 
For a description of {\tt MultiNest}, see Ref.~\cite{Feroz:2007kg,Feroz:2008xx}

When $P(\bTheta|\bD,H)$ is known, one can relatively easily obtain the posterior probability distribution, often just called {\it the posterior}, for some function $f(\bTheta)$ of the parameters, {\it e.g.}\ the predicted mass of a (new) particle in the model. In this work we will mainly study two-dimensional posterior distributions $P(X,Y|\bD,H)$ for pairs of observables $X$, $Y$. To obtain these we start from the joint posterior distribution for both the observables and the parameters, $P(X,Y,\bTheta|\bD,H)$. 

Since the two observables are functions $f_X$ and $f_Y$ of the model parameters, this joint posterior is related to the parameter posterior $P(\bTheta|\bD,H)$ simply by\footnote{Note that multiple parameter points $\bTheta_i$ may map to the same value of an observable.}
\begin{eqnarray}
  P(X,Y,\bTheta|\bD,H) &=& P(X,Y|\bTheta,\bD,H) P(\bTheta|\bD,H)
  \nonumber \\
  &=& \delta(f_X(\bTheta) - X) \delta(f_Y(\bTheta) -Y) P(\bTheta|\bD,H).
\end{eqnarray}
Therefore, given a set of parameter points $\bTheta_i$ drawn from $P(\bTheta|\bD,H)$, we simply calculate $f_X(\bTheta)$ and $f_Y(\bTheta)$ for each $\bTheta_i$ to obtain a set of samples $(X,Y,\bTheta)_i$ distributed according to $P(X,Y,\bTheta|\bD,H)$.

Finally, as the two-dimensional posterior of the observables $P(X,Y|\bD,H)$ is related to  $P(X,Y,\bTheta|\bD,H)$ through integration (so-called ``marginalization'') over the model parameters,
\begin{equation}
  P(X,Y|\bD,H) = \int P(X,Y,\bTheta|\bD,H)\, d\bTheta,
\end{equation}
an approximation to the two-dimensional posterior distribution $P(X,Y|\bD,H)$ that we want can be obtained by histogramming the Monte Carlo samples $(X,Y,\bTheta)_i$ in terms of the observables $X$ and $Y$.

\subsection{Scan set-up}
\label{sec:scan_set-up}

The scan uses {\tt MultiNest~2.17}~\cite{Feroz:2007kg,Feroz:2008xx} to explore the Natural SUSY parameter space based on a likelihood function $\llh(\bTheta)$ defined from observables and a choice of prior distribution function $\pi(\bTheta)$. In order to calculate the necessary observables for each sampled parameter point we make use of several public codes: the particle spectrum, including the effects discussed in Section \ref{sec:massdiff} for the chargino--neutralino mass difference, is calculated by {\tt SoftSusy 3.3.5}~\cite{softsusy}. Furthermore, Higgs masses are calculated by {\tt FeynHiggs 2.9.4}~\cite{feynhiggs1,feynhiggs2,feynhiggs3,feynhiggs4}, while B--physics observables are calculated using routines contained in {\tt MicrOMEGAS 2.4.5}~\cite{micromegas1,micromegas2,micromegas3}. The {\tt Python} SUSY Les Houches Accord (SLHA) interface {\tt PySLHA} is used to ease communication between the different codes, as well as processing the final results~\cite{Buckley:2013jua}. The scan output is a representative posterior sample of the parameter space with $\sim 130$k equally weighted points. This is achieved after combining the results from four independent scans, each running in parallell on 48 modern CPUs for 50 hours, using 6000 live points for the nested sampling. In total, the scan visits some 10M parameter points.

The prior distribution function $\pi(\bTheta)$ is constructed from the priors $\pi_i(\theta_i)$ for each individual parameter,
\begin{equation}
  \pi(\bTheta) = \prod\limits_i\pi_i(\theta_i).
\end{equation}
The authors of~\cite{Cabrera:2008tj} show that using logarithmically flat priors, $\pi_i(\theta_i) \propto \theta_i^{-1}$, corresponds to penalising unnatural models with exactly the Barbieri--Giudice fine-tuning measure~\cite{Barbieri:1987fn}
\begin{equation}
  c_{\theta_i}=\left|\frac{\partial\ln M_Z^2}{\partial\ln \theta_i }\right|,
  \label{eq:finetuning}
\end{equation}
for a parameter $\theta_i$ of the model. For this reason, we use log-priors for all dimensionful parameters of the model, while for $\tan\beta$ a linear prior is chosen. The SM parameters $m_t$, $m_b$, $M_Z$, $\alpha_{EM}$ and $\alpha_s$ are included as nuisance parameters with Gaussian priors.

For the parameter ranges, taking too wide intervals is problematic for scan
efficiency; nevertheless,  we want to ensure that we capture all of
the parameter space which can reasonably be termed natural.  From
arguments in the literature~\cite{Papucci:2011wy}, we consider stops
and sbottoms with masses below 1~TeV as plausible in a natural model,
thus this value is chosen as the upper bound for the respective soft masses. For the same reason the value of $|\mu|$  is also kept below 1~TeV, but $\mu$ can be negative. For the gluino an upper mass of 2~TeV is used since the naturalness bound is weaker. The other gaugino masses are allowed to range up to 5~TeV, and are allowed to be negative. As we have seen in Sec.~\ref{sec:mdiff}, it is essential to cover negative $M_1$ values and large values of $M_2$ in order to explore the most degenerate parts of the parameter space for higgsinos. Prior distributions and ranges for all scanned parameters are summarised in Table~\ref{tab:ScanPar}.

\begin{table}
  \begin{center}
    \begin{small}
      \begin{tabular}{lccc}
        \toprule
        Parameter          &      Range             & Prior       & Reference \\
        \midrule
        $M_1$              &  $[-5000,5000]$        &  log        &   -    \\
        $M_2$              &  $[-5000,5000]$        &  log        &   -    \\
        $M_3$              &  $[-2000,2000]$        &  log        &   -    \\
        $\mu$              &  $[-1000,1000]$        &  log        &   -    \\
        $m_{A^0}$          &  $[0,2000]$            &  log        &   -    \\
        $m_{Q_3}$          &  $[0,1000]$            &  log        &   -    \\
        $m_{u_3}$          &  $[0,1000]$            &  log        &   -    \\
        $m_{d_3}$          &  $[0,1000]$            &  log        &   -    \\
        $A_0$              &  $[-7000,7000]$        &  log        &   -    \\
        $\tan\beta$        &  $[2,60]$              &  linear     &   -    \\
        \midrule
        $m_t$              & $173.4 \pm 1.0$        &  Gaussian   & \cite{CMS:2012fya}  \\
        $m_b$              & $4.18  \pm 0.03$       &  Gaussian   & \cite{Beringer:1900zz}  \\
        $M_Z$              & $91.1876 \pm 0.0021$   &  Gaussian   & \cite{Beringer:1900zz}  \\
        $\alpha^{-1}$      & $127.944 \pm 0.014$    &  Gaussian   & \cite{Beringer:1900zz} \\
        $\alpha_s$         & $0.1184 \pm 0.0007$    &  Gaussian   & \cite{Beringer:1900zz} \\
        \bottomrule
      \end{tabular}
    \end{small}
\caption{List of scanned parameters with ranges and priors. All dimensionful parameters are given in GeV. In the case of log priors, the prior distributions are set to zero over the ranges $(-25,25)$ and $(0,25)$ GeV for signed and non-negative parameters, respectively.}\label{tab:ScanPar}
  \end{center}
\end{table}

As the basis for the likelihood $\llh(\bTheta)$ is the PDF of the observables and we assume the observables to be independent, the likelihood is constructed as
\begin{equation}
  \llh(\bTheta) = \prod\limits_i\llh_i(\bTheta),
\end{equation}
where $\llh_i(\bTheta)$ are the likelihoods for the individual observables. For every measured observable we use a gaussian likelihood in terms of the observable, except for BR$(B_s \rightarrow \mu\mu)$ for which we use the likelihood function published by the LHCb experiment~\cite{Aaij:2013aka,Aaij:2013aka_sup}.\footnote{As the experimental likelihood for $BR(B_s \rightarrow \mu\mu)$ published by LHCb covers a fairly wide range of $BR$ values~\cite{Aaij:2013aka,Aaij:2013aka_sup}, we use this for our scan, even though the corresponding measurement published by CMS is slightly more constraining~\cite{Chatrchyan:2013bka}.} For lower bounds on sparticle masses the likelihood is either 1 or 0 depending on whether the bound is satisfied or not.

The set of measured observables that enter in the complete scan
likelihood is listed in Table~\ref{tab:Constraints} together with the
choice of likelihood functional form, while the set of applied mass limits
is given in Table~\ref{tab:MassCuts}.
We use information on the
lightest Higgs mass from the LHC,\footnote{Note that by using this
  constraint we are demanding that the corrections responsible for the
  large measured Higgs mass come from the MSSM sector alone and not
  from {\it e.g.}\ extra NMSSM degrees of freedom.} the current average
experimental value of $M_W$ and a selection of the most constraining
B--physics results.
The reader may wonder why we do not consider the
measurement of $g-2$ for the muon. To approach the experimental value, large values of $\tan\beta$ are preferred and as discussed in Section~\ref{sec:mdiff} this will require even larger values of $M_2$ in order to give small or negative $\Delta m$ in the higgsino limit.\footnote{The leading chargino loop contribution to $g-2$ scales linearly with $\tan\beta$. In the higgsino limit the chargino coupling is proportional to the muon mass, thus satisfying $g-2$ is an issue.} Including the measurement can only further strengthen our conclusions. Dropping the measurement, in light of the lingering controversy over the difference between experiment and SM prediction, is thus a conservative approach here.

For the hard cuts a selection of direct and indirect mass limits from
LEP and Tevatron are used. Here we apply a conservative interpretation of published
limits at 95\% C.L., paying close attention to the region of validity
in parameter space; {\it e.g.}\ the absence of GUT-motivated relations
for the gaugino masses relaxes many of the standard bounds, and the
possibility of a higgsino LSP complicates matters further.
Assuming R-parity is conserved, we apply the most conservative of the MMCP limits in~\cite{Abazov:2011pf} on models where the lightest chargino is lighter than the lightest neutralino, which is the one for pure higgsinos. The impact of more constraining experimental searches for long-lived charged particles when $\Delta m > 0$ is discussed in Section~\ref{sec:imp}. Further details on the applied experimental mass limits are given in Table~\ref{tab:MassCuts}, together with the list of mass cuts employed to constrain the scan to the Natural SUSY scenario.

\begin{table}
  \begin{center}
    \begin{small}
      \begin{tabular}{lccc}
        \toprule
        Observable                     &     Constraint                       &  Likelihood              & Reference\\
        \midrule
        $M_W$                          &  $80.385 \pm 0.021$                  &  Gaussian                & \cite{Aaltonen:2013iut} \\ 
        $m_h$                          &  $125.7 \pm 2.0$                     &  Gaussian                & \cite{CMS:yva} \\  %
        $BR(B_s \rightarrow \mu\mu)$   &  $2.9^{+1.1}_{-1.0} \times 10^{-9}$  &  from experiment         & \cite{Aaij:2013aka,Aaij:2013aka_sup} \\
        $BR(b \rightarrow s \gamma)$   &  $(3.55 \pm 0.33) \times 10^{-4}$    &  Gaussian                & \cite{Amhis:2012bh} \\
        $R(B \rightarrow \tau \nu)$    &  $1.63 \pm 0.54$                     &  Gaussian                & \cite{Amhis:2012bh} \\
        \bottomrule
      \end{tabular}
    \end{small}
    \caption{List of the constraints entering the total scan likelihood function. All masses are given in GeV. Experimental and theoretical errors have been added in quadrature.}\label{tab:Constraints}
  \end{center}
\end{table}

\begin{table}
  \begin{center}
    \begin{small}
      \begin{tabular}{lccc}
        \toprule
        Observable                     &     Constraint       &  Conditions                    & Reference/Comment\\
        \midrule
        $m_{\widetilde{\chi}^\pm_1}$   &  $ > 217 $           &  $  \Delta m < 0$              & \cite{Abazov:2011pf}\\
        $m_{\widetilde{\chi}^\pm_1}$   &  $ > 102 $           &  $ 0 < \Delta m < m_\pi $      & \cite{Abdallah:2003xe} \\
        $m_{\widetilde{\chi}^\pm_1}$   &  $ > 70 $            &  $ m_\pi < \Delta m < 3 $, $|M_2| < |\mu|$  & \cite{Abdallah:2003xe}\\
        $m_{\widetilde{\chi}^\pm_1}$   &  $ > 75 $            &  $ m_\pi < \Delta m < 3 $, $|M_2| > |\mu|$  & \cite{Abdallah:2003xe}\\
        $m_{\widetilde{\chi}^\pm_1}$   &  $ > 97 $            &  $ 3 < \Delta m < 5 $          & \cite{Abdallah:2003xe}\\
        $m_{\widetilde{\chi}^\pm_1}$   &  $ > 102 $           &  $ 5 < \Delta m $              & \cite{Abdallah:2003xe}\\
        $m_{\tilde{g}}$                &  $ > 51 $            &  -                             & \cite{Kaplan:2008pt}\\
        $m_{\tilde{t}_1}$              &  $ > 63 $            &  -                             & \cite{Heister:2002hp}\\
        $m_{\tilde{t}_1}$              &  $ < 1000$           &  -                             & imposed for naturalness\\
        $m_{\tilde{t}_2}$              &  $ < 1000$           &  -                             & imposed for naturalness\\
        $m_{\tilde{b}_1}$              &  $ < 1000$           &  -                             & imposed for naturalness\\
        $m_{\tilde{g}}$                &  $ < 2000$           &  -                             & imposed for naturalness\\
        $\mathrm{min}(m_{\widetilde{\chi}^0_1},m_{\widetilde{\chi}^\pm_1})$
                                       &  $ < 500$  &  -   & imposed for naturalness\\
        \bottomrule
      \end{tabular}
    \end{small}
    \caption{List of hard mass cuts employed in the scan. All masses are given in GeV.}\label{tab:MassCuts}
  \end{center}
\end{table}

No dark matter constraints have been applied. For a chargino lighter than the lightest neutralino, the most natural dark matter candidate would be the gravitino. The possibility of the gravitino as the actual LSP would throw most such constraints out the window.\footnote{There could be new relevant constraints from the neutralino/chargino lifetime compared to Big Bang Nucleosynthesis that we do not consider here.}

Nor do we apply direct collider constraints from the LHC in the
scan. These often have a complicated dependence on the MSSM mass
spectrum, {\it e.g.}\ the many searches for jets and missing energy
depend on the first and second generation squark masses which we have
fixed. To avoid such constraints it suffices to have these squark
masses sufficiently large. Searches for EW production can likewise be
avoided with large slepton masses, since the alternative of direct
chargino production with short lived charginos is invisible due to the soft pion decay products in our scenarios.

Constraints from the production of third generation squarks or gluinos
have already been discussed in the literature~\cite{Papucci:2011wy,Buchmueller:2013exa,Kowalska:2013ica}.\footnote{The authors of~\cite{Kowalska:2013ica} also note the difficulties in having a small mass difference for higgsino LSPs.} While these now
narrow down the allowed parameter space of Natural SUSY, for the
purpose of this paper it is sufficient to observe that there are parts
of the parameter space that survive the current constraints.
For example, this is the case in scenarios with relatively high
third-generation squark masses, or scenarios with relatively small
third generation squark--higgsino mass
difference~\cite{Buchmueller:2013exa}. Such parameter choices do not
have a significant impact on our conclusions on the chargino--neutralino mass difference which is the focus here.

In view of the above, we focus on the predictions of indirect
constraints for the viable parameter space. However, we shall of course
discuss the impact of direct LHC searches for long lived charged
particles in
Section \ref{sec:imp} below.

\subsection{Results of scan}

In Fig.~\ref{fig:deltamvsmchi01} (left) we show our main result, the marginalised posterior probability distribution in the $m_{\tilde\chi_1^0}-\Delta m$ plane, for a wide range of mass differences (upper panel) and focused on small mass differences (lower panel). As expected from naturalness considerations, the lightest neutralino is fairly light, with the upper 68\% C.R.\  extending up to $\sim 400$~GeV.\footnote{There is a slight mismatch in the reach of the 68\% C.R.\ in the two panels due to the plot resolution.} The region of very light neutralinos in the upper panel may seem to be in conflict with the PDG limit of $m_{\tilde\chi_0^1} > 46$~GeV~\cite{Beringer:1900zz}. However, this bound assumes a GUT relation for the gaugino masses. Other limits, such as from the invisible $Z$-width, excludes very little of the parameter space not excluded by chargino bounds~\cite{Dreiner:2009yk}.

\begin{figure}[h!]
\begin{center}
  \includegraphics[width=0.49\textwidth]{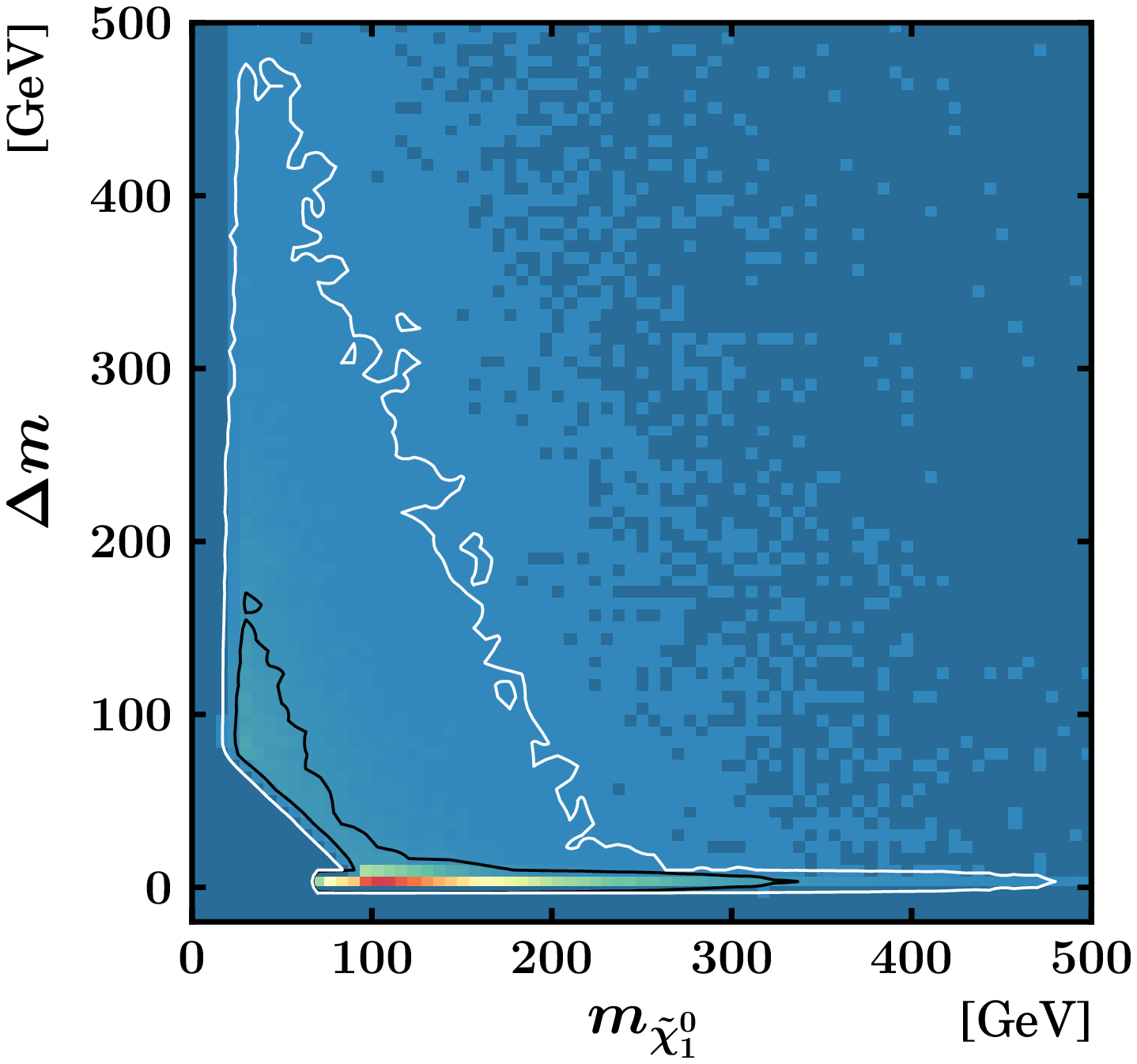}
  \includegraphics[width=0.49\textwidth]{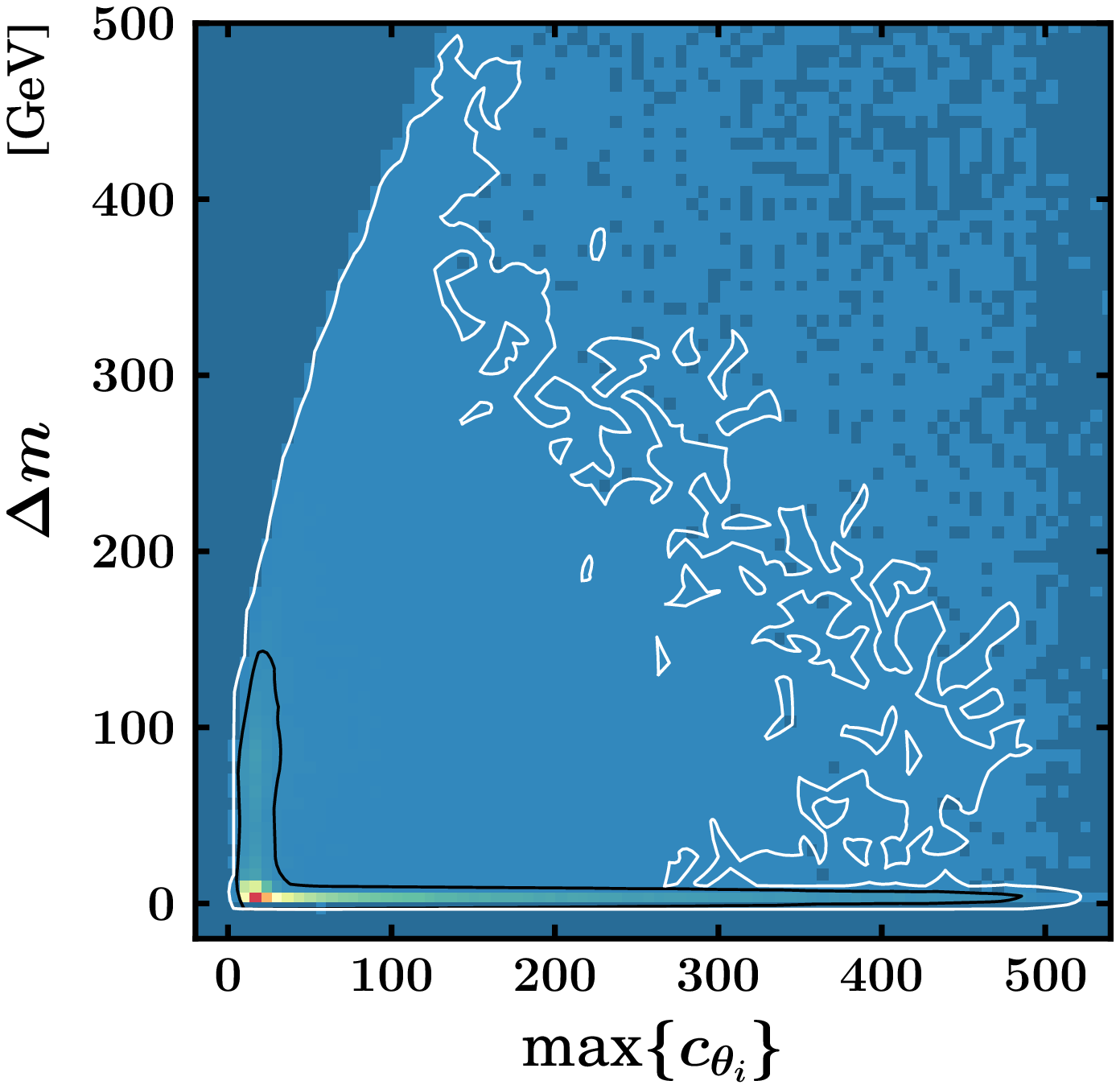}
\end{center}
\begin{center}
  \includegraphics[width=0.49\textwidth]{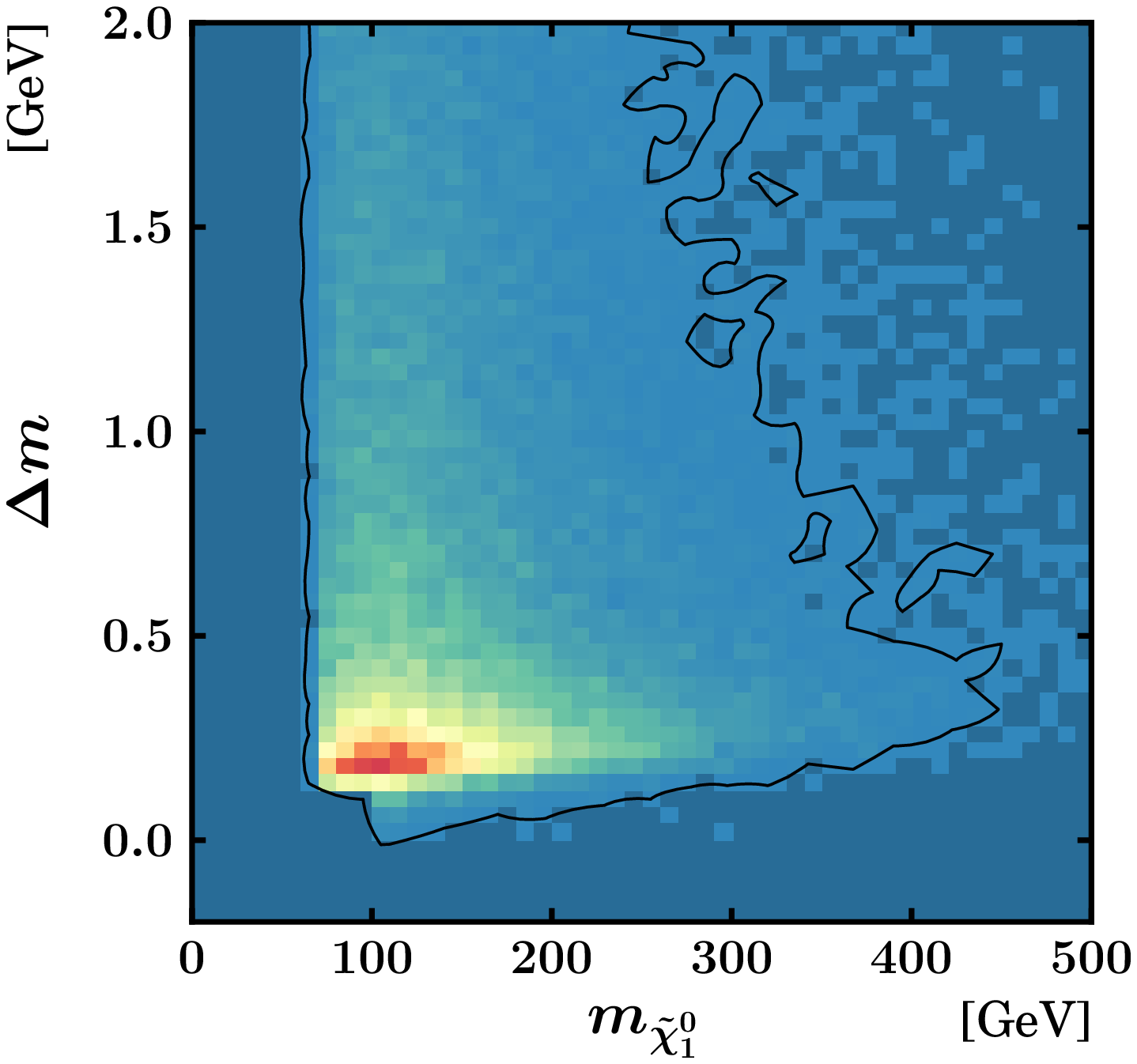}
  \includegraphics[width=0.49\textwidth]{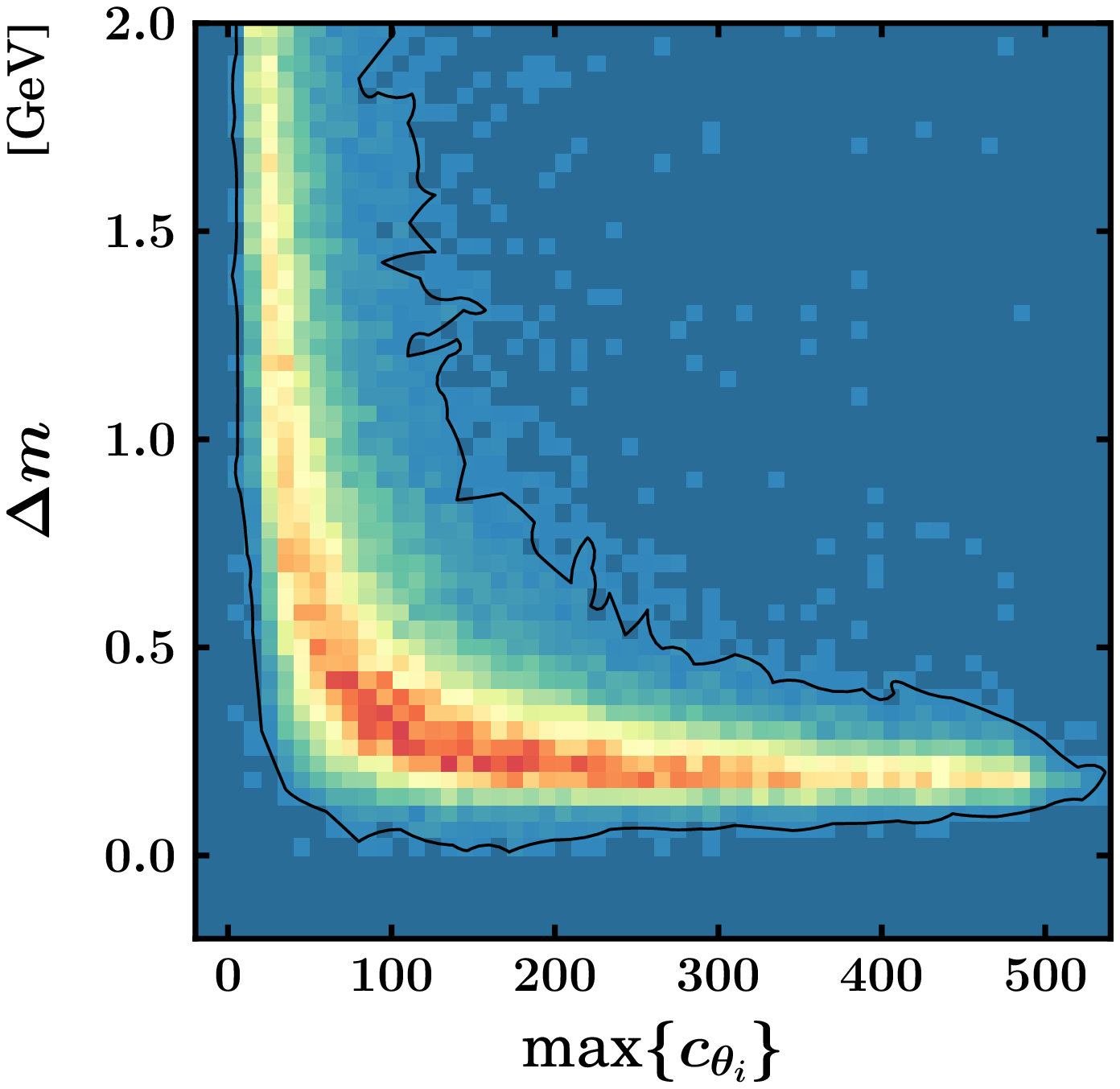}
\end{center}
\caption{Posterior distribution of $\Delta m$ versus the lightest neutralino mass $m_{\tilde\chi_0^1}$ (top left) and $\Delta m$ versus the fine-tuning measure $c\equiv \max\{c_{\theta_i}\}$ (top right), where $\{\theta_i\}$ is the set of free model parameters in the scan. The black and white lines show the 68\% and 95\% C.R.,\ respectively. The bottom panels depict the same distributions, but focused on small mass differences. }\label{fig:deltamvsmchi01}
\end{figure}

The negative mass differences that are in principle possible for higgsinos, see Sec.~\ref{sec:mdiff}, are outside the 95\% C.R. By removing the relevant constraints we have checked that the physical explanation of this traces back to a combination of three constraints: first, the relatively large Higgs mass prefers large stop masses and high $\tan\beta$ in order to get sufficient radiative corrections in the MSSM. As noted in Sec.~\ref{sec:mdiff}, a high $\tan\beta$ disfavours small chargino--neutralino mass differences due to the small numerical value of the factor $\sin{2\beta}$ in Eq.~(\ref{eq:dmhiggsino}), which gives the  mass difference for the higgsino case. This dependence in (\ref{eq:dmhiggsino}) stems from higgs-higgsino-gaugino terms in the Lagrangian, which enter into the tree-level neutralino and chargino mass matrices.

The second and third significant constraint is the LEP chargino mass bound and the Tevatron limit on MMCPs. These exclude, respectively, points with small but positive $\Delta m$ for chargino masses below $\sim 100$ GeV, and negative $\Delta m$ for chargino masses below $\sim 200$ GeV. In order to have small or negative $\Delta m$ when we have large $\tan\beta$ from the Higgs mass constraint we need, again from Eq.~(\ref{eq:dmhiggsino}) and see also Fig.~\ref{fig:mdiff}, that $\mu$ is small, and $M_1$ is relatively small and negative, of the order of $\sim 200$ GeV. This is then prevented by the LEP and Tevatron bounds, because this will result in too light charginos.

At small mass differences we observe a focusing in Fig.~\ref{fig:deltamvsmchi01} at around 150--200 MeV mass difference. This is due to a wino-like chargino and neutralino, $M_2<|\mu|$, where the mass difference comes from radiative corrections of order $\alpha M_W$, with some numerical prefactor. Again the Higgs mass constraint plays an important role: from Eq.~(\ref{eq:dmwinotree}) we see that large $\tan\beta$, which implies small $\sin{2\beta}$, significantly depletes the tree-level contribution from all terms of order less than $1/\mu^4$ in the expansion. The wino LSP dominates the smaller mass differences, as we discuss below.

The fine-tuning parameter of Eq.~(\ref{eq:finetuning}) versus the mass
difference is shown in Fig.~\ref{fig:deltamvsmchi01} (right). Because
of the naturalness inherent in the log-prior we see a rather low
fine-tuning  for significant fractions of the preferred parameter
space.
The banana of fine-tuning observed when focusing on low mass
differences in the lower panel is again an effect of dominantly wino
LSPs at small mass differences. When parameter combinations that have a wino LSP, $M_2<|\mu|, |M_1|$, are selected in the scan, then to achieve a mass difference below $\sim 0.5$~GeV,
the $|\mu|$ parameter must also be relatively large in order for the radiative correction to be
the dominant contribution to the mass difference, see Eq.~(\ref{eq:dmwinotree}). This in turn gives rise to additional fine-tuning, since in the MSSM, $\mu$ enters directly into the tree-level expression for $M_Z$ in Eq.~(\ref{eq:finetuning}).

\begin{figure}[h!]
\begin{center}
  \includegraphics[scale=0.46]{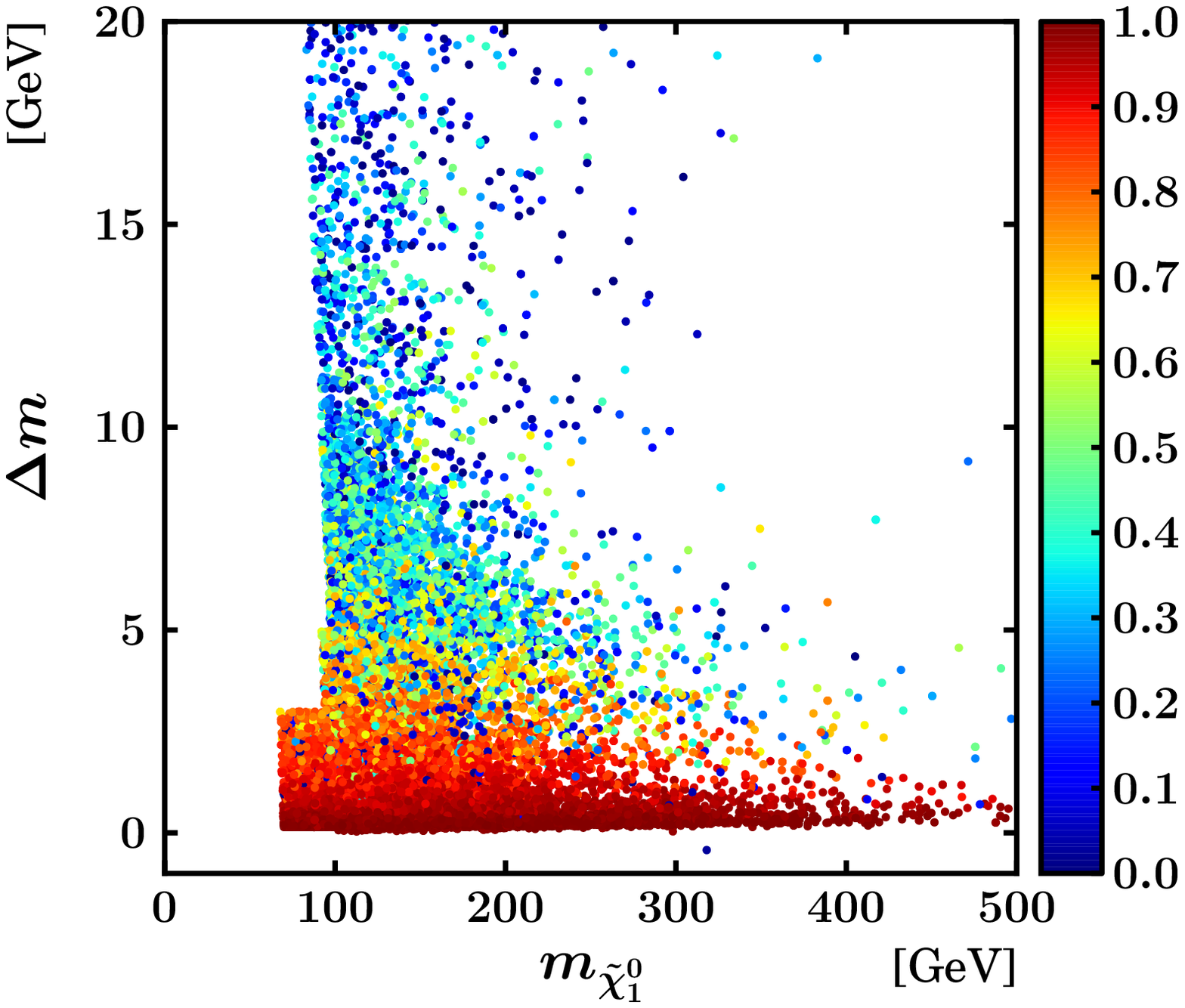}
  \includegraphics[scale=0.46]{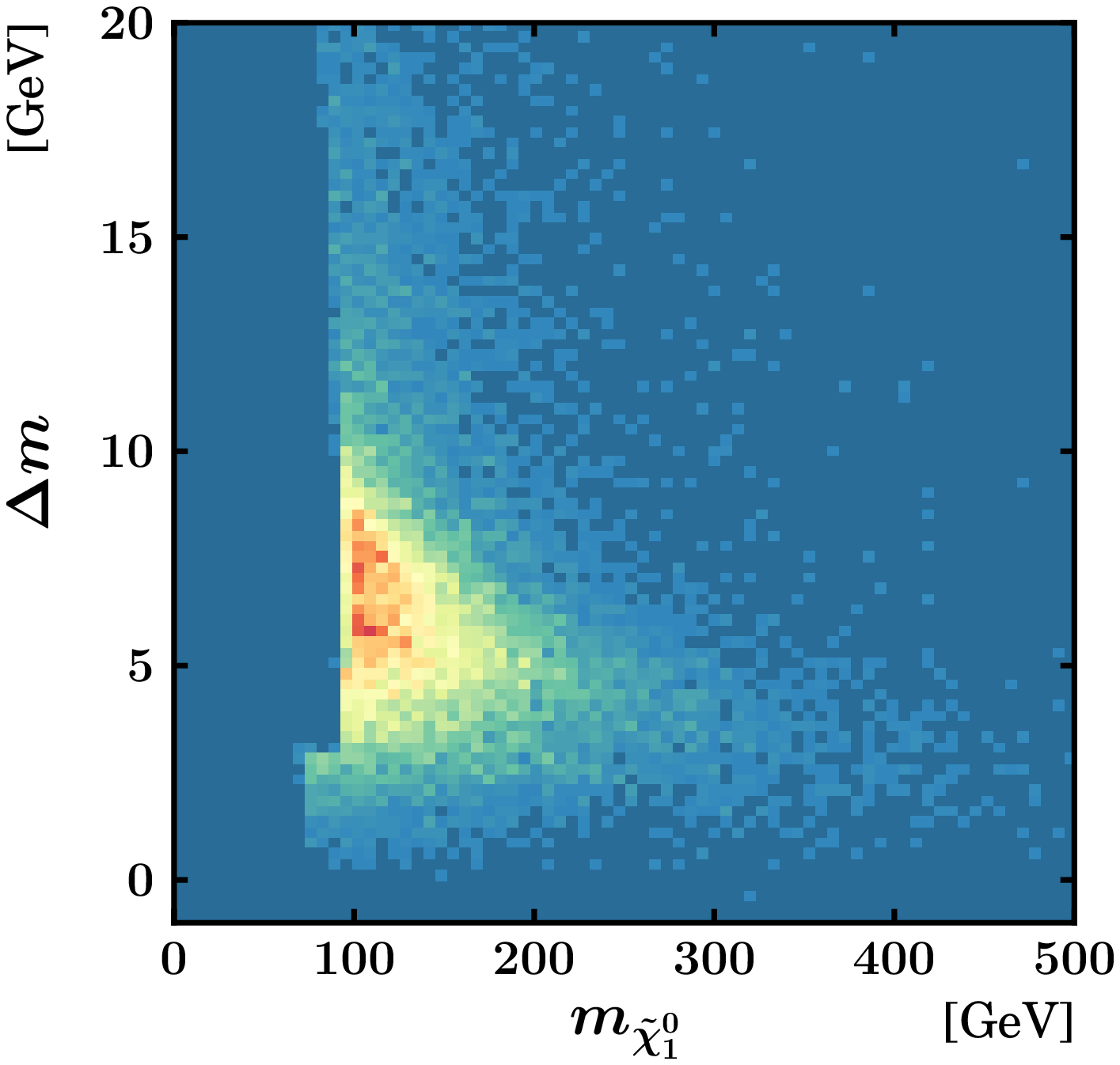}
\end{center}
\caption{Scatter plot of the wino component of the neutralino (left) and posterior distribution of $\Delta m$ versus the lightest neutralino mass $m_{\tilde\chi_0^1}$ for  a higgsino-like neutralino (right).}\label{fig:deltamvsmchi01winohiggsino}
\end{figure}

The composition of the neutralinos in these models is shown in Fig~\ref{fig:deltamvsmchi01winohiggsino}, where the left panel shows a scatter plot of the wino component. This demonstrates what we have claimed above: the smaller chargino--neutralino mass differences in Natural SUSY are dominated by winos, not higgsinos.

If we split the posterior parameter space into wino and higgsino
categories depending on the dominant component, the neutralino mass
versus $\Delta m$ distribution is shown in
Fig.~\ref{fig:deltamvsmchi01winohiggsino} (right) for a
dominantly higgsino $\tilde\chi_1^0$. This distribution is centered
around mass differences of the order of 5--10 GeV. There is a small
fraction of the parameter space with very small mass differences, but
this makes up an almost insignificant volume in the total space. This
is due to (i) the requirement of naturalness built into the
logarithmic priors, limiting high values of $|M_2|$, and (ii) to the
relatively high Higgs mass measured that favours large $\tan \beta$ (see discussion in Section~\ref{sec:mdiff}). We discuss the consequences of these results in the next section.

\section{Implications for collider searches}
\label{sec:imp}
For small mass differences in R-parity conserving models the relevant decay modes of the chargino are $\tilde\chi_1^\pm\to\tilde\chi_1^0(e^\pm\nu,\mu^\pm\nu)$ and $\tilde\chi_1^\pm\to\tilde\chi_1^0\pi^\pm$, where the latter is dominant down to threshold~\cite{Chen:1996ap}. Below this threshold the chargino lifetime increases significantly, and this is where effects from a long-lived chargino are expected.

\begin{figure}[h!]
\begin{center}
 \includegraphics[width=0.70\textwidth]{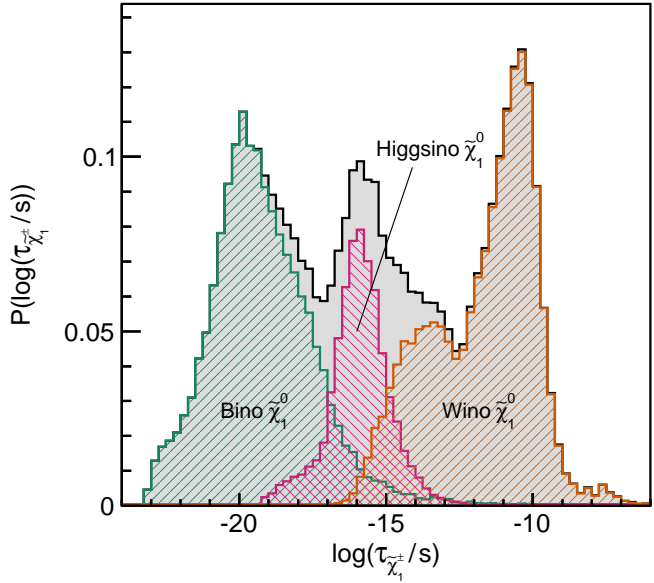}
\end{center}
\caption{Posterior distribution of the lifetime $\tau$ for
  $\tilde\chi_1^\pm$ (black).
The distributions for sub-samples with a bino (green), higgsino (magenta)
and wino (brown) dominated neutralino are also shown.}\label{fig:lifetime}
\end{figure}

In Fig.~\ref{fig:lifetime} we show the posterior lifetime distribution
for the chargino in our scan, also separated into models with bino,
wino or higgsino-dominated neutralinos. The longer lifetimes are
dominated by winos, with a distinct peak for the loop-dominated
$\Delta m$ around 150~MeV. The higgsinos  result to medium lifetimes,
mostly controlled by a typical 5--10~GeV
mass difference. This is to be compared to the shortest lifetimes and
much larger mass
differences of a bino LSP, with the chargino being either a wino or a higgsino.

Note that Fig.~\ref{fig:lifetime} carries another message: within Natural SUSY,
the relative probability for a bino, wino or higgsino LSP (the area below
the respective curves) is rather similar under the constraints applied,
with the higgsino scenario being slightly disfavoured.

The collider phenomenology of charginos at the LHC depends on the
lifetime shown in Fig.~\ref{fig:lifetime}. Typically, most charginos
will, due to their significant mass, be produced with momenta smaller than
their mass, but still at relatively large velocities because of the large
centre-of-mass energy of the proton collision and possible feed-down
energy from the decay of heavier sparticles. This means that $c\tau$
provides a good indication of the distance travelled by these
charginos before decay.\footnote{Even in the extreme case where a single
  100 GeV chargino picks up a 1 TeV boost the $\gamma$-factor gives a
  factor of 10 on this distance.} The phenomenology can be classified
as in \cite{Barr:2002ex}, where values of $c\tau\gsim 1$~cm (or
$\tau\gsim 3.3\times10^{-11}$~s) give a substantial number of kinked
tracks in the inner detector of ATLAS.

For even smaller mass differences below the pion threshold, giving $c\tau\gsim1$~m, a significant number of charginos can be obtained in the muon system as well. At higgsino (wino) masses below 217~GeV (267~GeV) these are already excluded by limits on metastable particles from the Tevatron~\cite{Abazov:2011pf}.

The kinked track scenarios should be easily detectable and have been looked for in current ATLAS and CMS searches which reach down to lifetimes of 0.06 ns ($c\tau=1.8$~cm)~\cite{Aad:2013yna}.
The search in \cite{Aad:2013yna} is interpreted in terms of a
minimal AMSB model with $m_0=1$~TeV, $\tan\beta = 5$ and $\mu>0$, and
a free gravitino mass $m_{3/2}$ that controls the wino LSP
mass. Despite this model dependence, since only direct chargino
production was considered, the limit set as a function of the chargino lifetime and mass, should apply fairly generally.\footnote{The limit is somewhat weaker for higgsinos because of a lower cross section~\cite{Raklev:2009mg}.}

\begin{figure}[h!]
\begin{center}
 \includegraphics[trim=0in 0in 0.5in 0in, clip=true, width=0.49\textwidth]{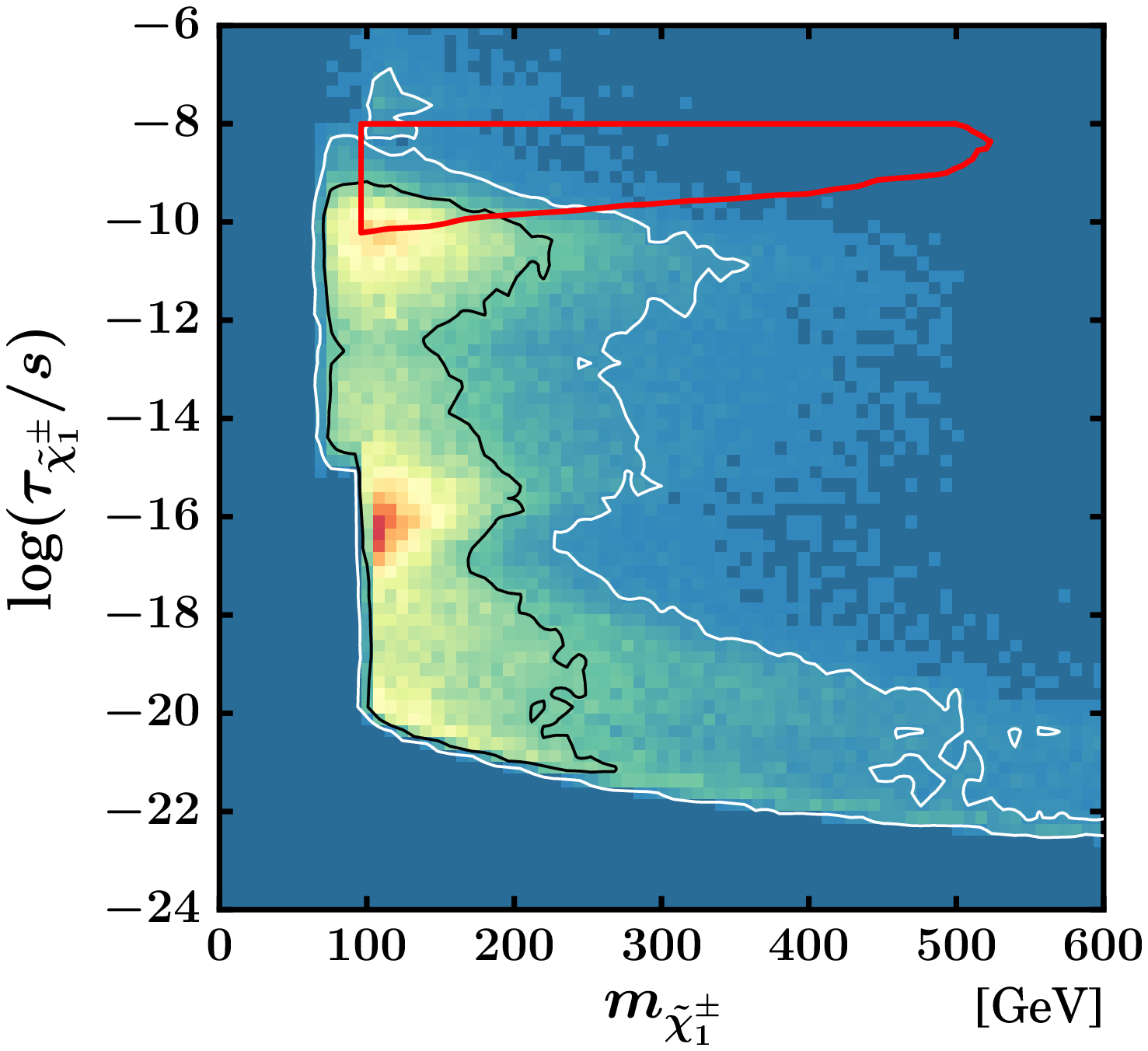}
 \includegraphics[trim=0in 0in 0.5in 0in, clip=true, width=0.49\textwidth]{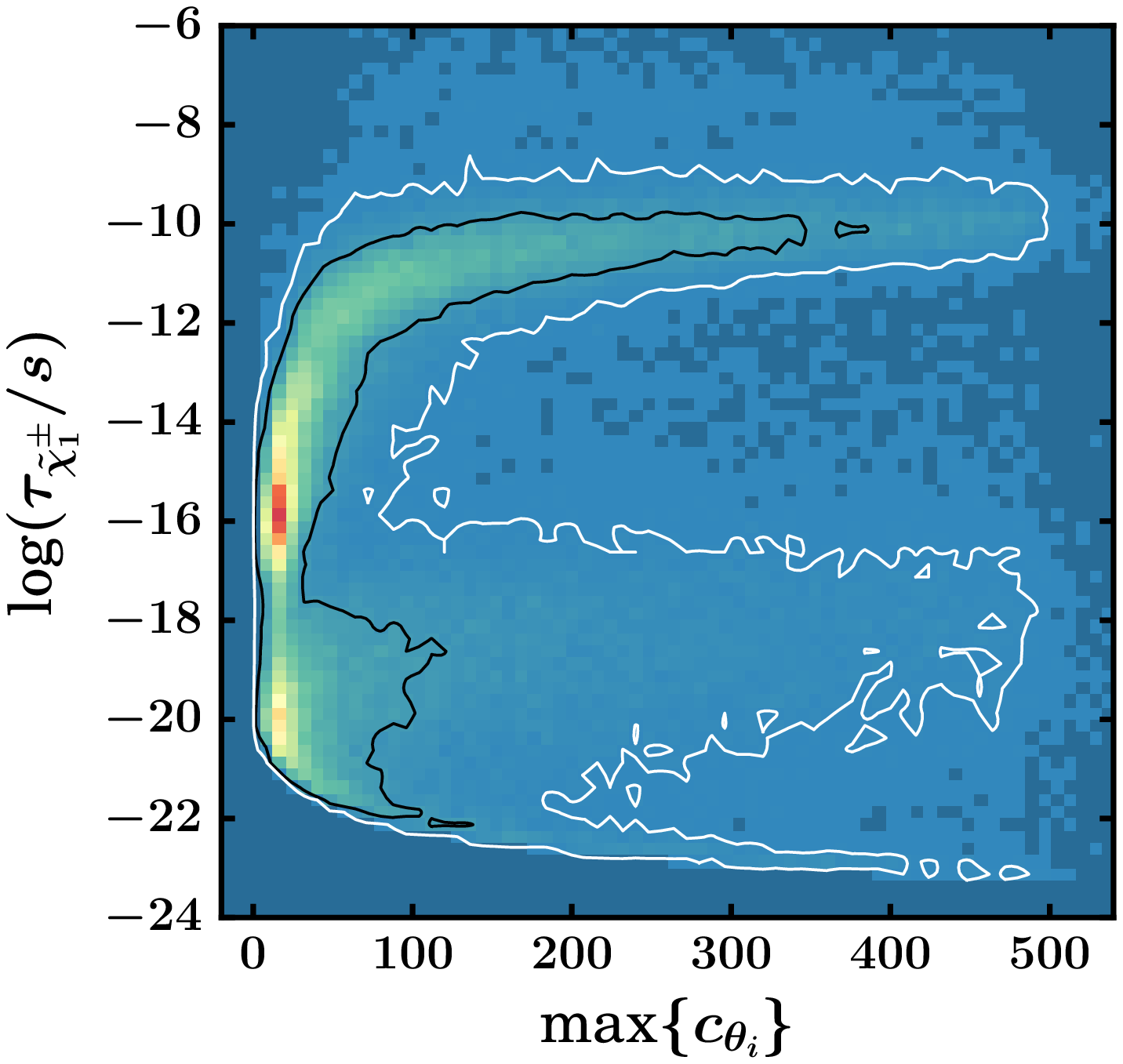}
\end{center}
\caption{Posterior distribution of chargino mass versus lifetime (left) and fine-tuning measure versus lifetime (right). The black and white lines show the 68\% and 95\% C.R.,\ respectively. Overlayed in red is the limit from a recent ATLAS search for kinked tracks~\cite{Aad:2013yna}.}\label{fig:mass_lifetime}
\end{figure}

The impact of this limit on our posterior distribution is shown in red in
Fig.~\ref{fig:mass_lifetime} (left). We see that the searches are
sensitive to, and have excluded, some models at the edge of the
Natural SUSY parameter space, but only for the wino LSP scenario. 
For the parameter space with lifetimes in the range $\sim10^{-10}$--$10^{-8}$~s and masses beyond the current reach of the LHC, the non-excluded region is mostly outside the 95\% C.R.

Only a very small volume of the preferred parameter space features actually long lived charginos with $\tau\gsim 10^{-8}$~s. For these models the Tevatron limits in~\cite{Abazov:2011pf} should apply, although making an accurate assessment in the mass reach is made difficult by a transitional interval in lifetime up to $\tau\sim 10^{-7}$~s, where only some charginos make it through the muon system. There is also a potentially interesting area left for long-lived winos between the LEP and LHC limits, from 70 to 100 GeV. However, a more detailed and less conservative implementation of LEP limits, or an extension of the ATLAS search to lower masses, would probably close this gap.

In Fig.~\ref{fig:mass_lifetime} (right) we see that the excluded part of
the parameter space with a chargino lifetime around $10^{-10}$~s is
also the most fine-tuned and least natural one,
for reasons that were discussed in relation to the fine-tuning banana in the previous section.

For lower lifetimes, down to values of $c\tau$ of order 30 $\mu$m,
chargino decays may potentially be observed as tracks with non-zero
impact parameter. However, due to large backgrounds from {\it e.g.}\
heavy quark decays it will be very difficult to trigger and identify SUSY events on this basis alone.

For Natural SUSY models with a higgsino LSP we saw in
Fig.~\ref{fig:deltamvsmchi01winohiggsino} that
a typical chargino--neutralino mass difference was in the region of 5--10~GeV (and $c\tau\sim 10^{-8}$~m). In this case the dominant chargino decay is not to a single pion, but instead to hadron showers (${\rm BR}\sim 70$\%) with a significant admixture of $l\nu$ ($\sim 30$\%). Both of these signatures are challenging to detect due to the relatively soft nature of the particles produced, but not impossible.

\section{Conclusions}
\label{sec:conclusions}
\setcounter{equation}{0}

Using a Bayesian parameter scan, we have explored within the MSSM framework to what extent Natural SUSY prefers long-lived charginos, and if it could even harbour a chargino lighter than the lightest neutralino, when confronted with a relatively conservative set of current direct and indirect constraints. We have also looked at the likely composition of the LSP given the same constraints.

We have shown that the bulk of the posterior probability has chargino--neutralino mass differences above 150--200 MeV, and that negative mass differences are outside the 95\% credible region of Natural SUSY in the MSSM. We note that the latter is a highly non-trivial consequence of the constraints applied, in particular the combination of the LHC Higgs mass measurement and a careful conservative consideration of the LEP and Tevatron constraints on the chargino.

As this preferred mass difference is larger than the pion mass threshold for the decay $\tilde\chi_1^\pm\to\tilde\chi_1^0\pi^\pm$, we find a corresponding preference for chargino lifetimes below $\sim 10^{-10}$~s, meaning that long-lived charginos is not an expected signature of Natural SUSY. The composition of the LSP is also fairly democratically distributed between bino, wino and higgsino, however, the addition of a dark matter relic density constraint, thus assuming that the lightest neutralino constitutes dark matter, is likely to change this.

In the region of preferred parameter space with the longest chargino
lifetimes, the mass degeneracy is due to wino dominance. In this
scenario, the level of degeneracy and thus the chargino lifetime
increases for larger values of $|\mu|$, until it is dominated by
loop-corrections.
As a result, the parameter region with the longest lifetimes is also a region with relatively high fine-tuning. Current experimental searches by ATLAS and CMS for long-lived particles are starting to exclude this scenario. Should such a long-lived particle be found just beyond current bounds, the conclusion, on the basis of the current paper, must be that it is either a wino, or that Natural SUSY is not realized.

A more natural scenario within Natural SUSY, where the lightest neutralino and chargino are predominantely higgsinos, favours chargino lifetimes below $\sim 10^{-15}$~s, and correspondingly chargino--neutralino mass differences above $\sim 5$~GeV. This scenario is challenging for discovery, but because of the relatively substantial mass difference, compared to the wino case, not impossible, and deserves serious experimental investigation.

\vspace{0.4 cm}

\noindent
{\bf Acknowledgements}\\
The CPU intensive parts of this work was performed on the Abel Cluster, owned by the University of Oslo
and the Norwegian metacenter for High Performance Computing (NOTUR), and operated by the Research Computing Services group at USIT, the University of Oslo IT-department. The computing time was given by NOTUR allocation NN9284K, financed through the Research Council of Norway. N.-E. Bomark is funded in part by the Welcome Programme of the Foundation for Polish Science.



\begin{thebibliography}{99}

\bibitem{Randall:1998uk}
  L.~Randall and R.~Sundrum,
  Nucl.\ Phys.\ B {\bf 557} (1999) 79
  [hep-th/9810155].

\bibitem{Giudice:1998xp}
  G.~F.~Giudice, M.~A.~Luty, H.~Murayama and R.~Rattazzi,
  JHEP {\bf 9812} (1998) 027
  [hep-ph/9810442].

\bibitem{Brust:2011tb}
  C.~Brust, A.~Katz, S.~Lawrence and R.~Sundrum,
  JHEP {\bf 1203} (2012) 103
  [arXiv:1110.6670 [hep-ph]].

\bibitem{Papucci:2011wy}
  M.~Papucci, J.~T.~Ruderman and A.~Weiler,
  JHEP {\bf 1209} (2012) 035
  [arXiv:1110.6926 [hep-ph]].


\bibitem{Chen:1996ap}
  C.~H.~Chen, M.~Drees and J.~F.~Gunion,
  Phys.\ Rev.\ D {\bf 55} (1997) 330
   [Erratum-ibid.\ D {\bf 60} (1999) 039901]
  [hep-ph/9607421].

\bibitem{Feng:1999fu}
  J.~L.~Feng, T.~Moroi, L.~Randall, M.~Strassler and S.~-f.~Su,
  Phys.\ Rev.\ Lett.\  {\bf 83} (1999) 1731
  [hep-ph/9904250].

\bibitem{Gherghetta:1999sw}
  T.~Gherghetta, G.~F.~Giudice and J.~D.~Wells,
  Nucl.\ Phys.\ B {\bf 559} (1999) 27
  [hep-ph/9904378].

\bibitem{Fairbairn:2006gg}
  M.~Fairbairn, A.~C.~Kraan, D.~A.~Milstead, T.~Sjostrand, P.~Z.~Skands and T.~Sloan,
  Phys.\ Rept.\  {\bf 438} (2007) 1
  [hep-ph/0611040].

\bibitem{Raklev:2009mg}
  A.~R.~Raklev,
  Mod.\ Phys.\ Lett.\ A {\bf 24} (2009) 1955
  [arXiv:0908.0315 [hep-ph]].

\bibitem{ATLAS-CONF-2013-058}
  G.~Aad {\it et al.}  [ATLAS Collaboration],
  ATLAS-CONF-2013-058.

\bibitem{Aad:2013yna}
  G.~Aad {\it et al.}  [ATLAS Collaboration],
  Phys.\ Rev.\ D {\bf 88} (2013) 112006
  [arXiv:1310.3675 [hep-ex]].
  


\bibitem{Chatrchyan:2013oca}
  S.~Chatrchyan {\it et al.}  [CMS Collaboration],
  JHEP {\bf 1307} (2013) 122
  [arXiv:1305.0491 [hep-ex]].


\bibitem{Kribs:2008hq}
  G.~D.~Kribs, A.~Martin and T.~S.~Roy,
  JHEP {\bf 0901} (2009) 023
  [arXiv:0807.4936 [hep-ph]].

\bibitem{Aad:2012tfa}
  G.~Aad {\it et al.}  [ATLAS Collaboration],
  Phys.\ Lett.\ B {\bf 716} (2012) 1
  [arXiv:1207.7214 [hep-ex]].
  

\bibitem{Chatrchyan:2012ufa}
  S.~Chatrchyan {\it et al.}  [CMS Collaboration],
  Phys.\ Lett.\ B {\bf 716} (2012) 30
  [arXiv:1207.7235 [hep-ex]].
  

\bibitem{Feroz:2007kg}
  F.~Feroz and M.~P.~Hobson,
  Mon.\ Not.\ Roy.\ Astron.\ Soc.\  {\bf 384} (2008) 449
  [arXiv:0704.3704 [astro-ph]].

\bibitem{Feroz:2008xx}
  F.~Feroz, M.~P.~Hobson and M.~Bridges,
  Mon.\ Not.\ Roy.\ Astron.\ Soc.\  {\bf 398} (2009) 1601
  [arXiv:0809.3437 [astro-ph]].

\bibitem{Martin:1997ns}
  S.~P.~Martin,
  In *Kane, G.L. (ed.): Perspectives on supersymmetry II* 1-153
  [hep-ph/9709356].

\bibitem{Feng:2008cn}
  T.~-F.~Feng, L.~Sun and X.~-Y.~Yang,
  Nucl.\ Phys.\ B {\bf 800}, 221 (2008)
  [arXiv:0805.1122 [hep-ph]].

\bibitem{Ellis:2008zy}
  J.~R.~Ellis, J.~S.~Lee and A.~Pilaftsis,
  JHEP {\bf 0810}, 049 (2008)
  [arXiv:0808.1819 [hep-ph]].

\bibitem{Cheung:2009fc}
  K.~Cheung, O.~C.~W.~Kong and J.~S.~Lee,
  JHEP {\bf 0906}, 020 (2009)
  [arXiv:0904.4352 [hep-ph]].

\bibitem{Altmannshofer:2009ne}
  W.~Altmannshofer, A.~J.~Buras, S.~Gori, P.~Paradisi and D.~M.~Straub,
  Nucl.\ Phys.\ B {\bf 830}, 17 (2010)
  [arXiv:0909.1333 [hep-ph]].

\bibitem{Giudice:1995qk}
  G.~F.~Giudice and A.~Pomarol,
  Phys.\ Lett.\ B {\bf 372} (1996) 253
  [hep-ph/9512337].

\bibitem{Cabrera:2008tj}
  M.~E.~Cabrera, J.~A.~Casas and R.~Ruiz de Austri,
  JHEP {\bf 0903} (2009) 075
  [arXiv:0812.0536 [hep-ph]].

\bibitem{Barbieri:1987fn}
  R.~Barbieri and G.~F.~Giudice,
  Nucl.\ Phys.\ B {\bf 306} (1988) 63.

\bibitem{CMS:2012fya}
  [CMS Collaboration],
  CMS-PAS-TOP-11-018.

\bibitem{Beringer:1900zz}
  J.~Beringer {\it et al.}  [Particle Data Group Collaboration],
  Phys.\ Rev.\ D {\bf 86}, 010001 (2012).

\bibitem{softsusy}
  B.C.~Allanach,
  Comput. \ Phys. \ Commun. \ {\bf 143} (2002) 305
  [arXiv:hep-ph/0104145].

\bibitem{feynhiggs1}
  S.~Heinemeyer, W.~Hollik and G.~Weiglein,
  Comput.\ Phys.\ Commun.\  {\bf 124} (2000) 76
  [hep-ph/9812320].

\bibitem{feynhiggs2}
  S.~Heinemeyer, W.~Hollik and G.~Weiglein,
  Eur.\ Phys.\ J.\ C {\bf 9} (1999) 343
  [hep-ph/9812472].

\bibitem{feynhiggs3}
  G.~Degrassi, S.~Heinemeyer, W.~Hollik, P.~Slavich and G.~Weiglein,
  Eur.\ Phys.\ J.\ C {\bf 28} (2003) 133
  [hep-ph/0212020].

\bibitem{feynhiggs4}
  M.~Frank, T.~Hahn, S.~Heinemeyer, W.~Hollik, H.~Rzehak and G.~Weiglein,
  JHEP {\bf 0702} (2007) 047
  [hep-ph/0611326].

\bibitem{micromegas1}
  G.~Belanger, F.~Boudjema, A.~Pukhov and A.~Semenov,
  Comput.\ Phys.\ Commun.\  {\bf 149} (2002) 103
  [hep-ph/0112278].

\bibitem{micromegas2}
  G.~Belanger, F.~Boudjema, A.~Pukhov and A.~Semenov,
  Comput.\ Phys.\ Commun.\  {\bf 174} (2006) 577
  [hep-ph/0405253].

\bibitem{micromegas3}
  G.~Belanger, F.~Boudjema, P.~Brun, A.~Pukhov, S.~Rosier-Lees, P.~Salati and A.~Semenov,
  Comput.\ Phys.\ Commun.\  {\bf 182} (2011) 842
  [arXiv:1004.1092 [hep-ph]].

\bibitem{Buckley:2013jua}
  A.~Buckley,
  arXiv:1305.4194 [hep-ph].

\bibitem{Aaij:2013aka}
  R.~Aaij {\it et al.}  [LHCb Collaboration],
  Phys.\ Rev.\ Lett.\  {\bf 111} (2013) 101805
  [arXiv:1307.5024 [hep-ex]].

\bibitem{Aaij:2013aka_sup}
  R.~Aaij {\it et al.}  [LHCb Collaboration], LHCB-PAPER-2013-046 (supplementary material), \url{http://cds.cern.ch/record/1563073/files/}.

\bibitem{Chatrchyan:2013bka}
  S.~Chatrchyan {\it et al.}  [CMS Collaboration],
  Phys.\ Rev.\ Lett.\  {\bf 111} (2013) 101804
  [arXiv:1307.5025 [hep-ex]].

\bibitem{Abazov:2011pf}
  V.~M.~Abazov {\it et al.}  [D0 Collaboration],
  Phys.\ Rev.\ Lett.\  {\bf 108} (2012) 121802
  [arXiv:1110.3302 [hep-ex]].

\bibitem{Aaltonen:2013iut}
  T.~A.~Aaltonen {\it et al.}  [CDF and D0 Collaborations],
  Phys.\ Rev.\ D {\bf 88} (2013) 5,  052018
  [arXiv:1307.7627 [hep-ex]].
  

\bibitem{CMS:yva}
  [CMS Collaboration],
  CMS-PAS-HIG-13-005.

\bibitem{Amhis:2012bh}
  Y.~Amhis {\it et al.}  [Heavy Flavor Averaging Group Collaboration],
  arXiv:1207.1158 [hep-ex].

\bibitem{Abdallah:2003xe}
  J.~Abdallah {\it et al.}  [DELPHI Collaboration],
  Eur.\ Phys.\ J.\ C {\bf 31} (2003) 421
  [hep-ex/0311019].

\bibitem{Kaplan:2008pt}
  D.~E.~Kaplan and M.~D.~Schwartz,
  Phys.\ Rev.\ Lett.\  {\bf 101} (2008) 022002
  [arXiv:0804.2477 [hep-ph]].

\bibitem{Heister:2002hp}
  A.~Heister {\it et al.}  [ALEPH Collaboration],
  Phys.\ Lett.\ B {\bf 537} (2002) 5
  [hep-ex/0204036].

\bibitem{Buchmueller:2013exa}
  O.~Buchmueller and J.~Marrouche,
  Int.\ J.\ Mod.\ Phys.\ A {\bf 29} (2014) 1450032
  [arXiv:1304.2185 [hep-ph]].
  

\bibitem{Kowalska:2013ica}
  K.~Kowalska and E.~M.~Sessolo,
  Phys.\ Rev.\ D {\bf 88} (2013) 075001
  [arXiv:1307.5790 [hep-ph]].

\bibitem{Dreiner:2009yk}
  H.~K.~Dreiner,
  AIP Conf.\ Proc.\  {\bf 1200} (2010) 73
  [arXiv:0910.1509 [hep-ph]].

\bibitem{Barr:2002ex}
  A.~J.~Barr, C.~G.~Lester, M.~A.~Parker, B.~C.~Allanach and P.~Richardson,
  JHEP {\bf 0303} (2003) 045
  [hep-ph/0208214].

\end{thebibliography}
\end{document}